# Mesosphere Light Scattering Depolarization
# During the Perseids Activity Epoch by WAPC Measurements

Oleg S. Ugolnikov*[1], Igor A. Maslov[1,2]

[1]Space Research Institute, Russian Academy of Sciences,
Profsoyuznaya st., 84/32, Moscow, 117997, Russia
[2]Moscow State University, Sternberg Astronomical Institute,
Universitetskii prosp., 13, Moscow, 119991, Russia
ougolnikov@gmail.com, imaslov@iki.rssi.ru.
*Corresponding author, phone: +7-916-391-73-00, fax: +7-495-333-51-78.

**Abstract**
The paper describes the study of field of scattered radiation in the mesosphere basing on wide-angle polarization camera (WAPC) measurements of the twilight sky background. Mid-August observations in 2012 and 2013 show the decrease of single scattering polarization value that can be related with Perseids meteor dust moderation in the upper mesosphere. Effect correlates with activity of tiny fraction of Perseids shower. Polarization and temperature analysis allows estimating the altitude of dust layer and character polarization of dust scattering.

**Keywords:** twilight sky background; mesosphere; polarization; meteor dust.

**1. Introduction.**

Mesosphere of the Earth is the moderation layer for interplanetary bodies with different sizes and masses. Larger meteoroids are burned creating the optical meteor events at the altitudes 60-100 km, smaller particles contribute to the upper atmospheric dust layer. The value of dust inflow is about several hundred tons per day. Meteor particles together with burning products have an influence to the mesosphere chemistry and optics. Space-originated dust is playing the role of condensation nuclei for polar mesospheric or noctilucent clouds (NLC).

Twilight sky method was considered as an effective tool for atmospheric scattering research for a long time. Elevation of the Earth's shadow helps to fix atmospheric layer with altitude depending on solar zenith angle and the sky point position. The basic advantage of the method is the possibility to measure the scattering matrix components for different scattering angles simultaneously. It is important for particles study, but hard to do by lidar or space limb measurements. However, this became possible just now using high-sensitive all-sky CCD-cameras.

Meteor dust can cause additional scattering, that will affect the dark twilight sky characteristics (Link and Robley, 1971). But we have to realize that this will be just the admixture to the single scattering field, which itself is the admixture to the total background dominated by multiple scattering. Date-to-date variation of basic fractions can exceed the total contribution of cosmic dust effect.

The problem of multiple scattering was being understood from the early XX century, but the theoretical and experimental methods of its separation (Rozenberg, 1966, Lebedinets, 1981) led to underestimated value of multiple scattering contribution. This meant to higher upper limit of twilight method study (up to 150-200 km) but unbelievable amount of dust contamination in mesosphere and even thermosphere.

Polarization analysis of twilight background is effective improvement of the method (Fesenkov, 1966) owing to less polarization of dust scattering compared the Rayliegh and even multiple scattering. It was used to find the possible effect of meteoric dust in upper atmosphere as early as in 1960s (Steinhorst, 1971), but actually it was below noise level (contributed not only by measurements, but also by multiple scattering, as it can be seen in the resulting graphs). Recent analysis shows that meteoric dust can significantly affect the general properties of the twilight sky only after the strong outbursts like Leonids in 2002 (Ugolnikov and Maslov, 2007), polarization turns out to be more sensitive than intensity.

To detect weaker dust effects, the method of single and multiple scattering separation must be used. Experimental one seems to be more preferable since high-order scattering in the lower atmosphere depends on variable troposphere conditions and is hard for exact computer modeling. Frequent polarization data in solar vertical points is found to be useful for this problem (Ugolnikov and Maslov, 2013a). The procedure described in that paper was then improved and expanded to the most part of celestial semi-sphere (Ugolnikov and Maslov, 2013b). It was turned out that the single scattering remains noticeable at altitudes below 90 km, this is the upper boundary layer for the analysis. Despite of low intensity contribution to the total background in the upper mesosphere, the altitude dependence and the polarization of single scattering can be found. These are close to Rayleigh values for most of twilights, that helps to find Boltzmann temperature distribution in the mesosphere. The results are in good agreement with present satellite data by TIMED/SABER (Russell et al., 1999) and EOS Aura/MLS (Schwartz et al., 2008) experiments and can be a less-expensive supplement to mesospheric temperature monitoring actual in the present time.

Small depolarization of single scattering was often observed and can be explained by sporadic dust inflow. The effect increased in first half of August of 2012. It was assumed that the reason was related with Perseids meteor shower. However, the number of observations in August, 2012 was not so high to confirm or refute it. The aim of this work is to clarify this using the data of the same month in 2013.

**2. Observations.**

The twilight sky measurements are hold in central Russia, 55.2°N, 37.5°E, about 60 km southwards from Moscow. The description of wide-angle polarization camera (WAPC) with device photo is performed in (Ugolnikov and Maslov, 2013b), it was not sufficiently changed in 2013. The camera operates in spectral band with effective wavelength equal to 540 nm, measuring the twilight sky intensity and polarization in different points of the wide sky area up to 70° around the zenith from the daytime till the deep night. Stars astrometry and photometry in the night sky images fix the camera orientation, field curvature and self polarization. It is also used to measure the multiplication of camera flat field and atmospheric transparency, separately for each night.

Perseids observation conditions in 2013 were optimal for northern hemisphere. As it was shown in (Ugolnikov and Maslov, 2013b), crescent Moon (phase below 0.5) does not cause the problems for data procession. Visual maximum of the shower (August, 12$^{th}$) followed the New Moon, and dark twilights, both evening and morning, were not affected by the scattered moonlight during pre-maximum and maximum epochs. Morning twilights in 2013 as the evening ones in 2012 were also moonlight-free during maximum and post-maximum periods.

**3. Single scattering analysis.**

Twilight sky data is being processed by the method described in (Ugolnikov and Maslov, 2013b). Effective single light scattering takes place in the mesosphere during the dark stage of twilight, at solar zenith angles 98-99°. The basic problem is high contribution of multiple scattering which becomes the dominant background component during this part of twilight. The method allows to fix the single scattering in the dawn area, where the Earth's shadow is lower than in opposite part of the sky. Using the data aside from solar vertical, the range of single scattering angles can be expanded, but it does not exceed several dozens of degrees.

The result of the routine is the intensity and polarization values of single scattering in the atmospheric column above the Earth's shadow as a function of the scattering angle $\theta$ (corresponding solar zenith angle for definite shadow altitude is a function of sky point). The character baseline altitude $h_B$ of this column is higher than the geometrical shadow altitude due to significant extinction of tangent solar emission in the lower atmosphere (see Fig.1). For the spectral band used in this paper, the difference is about 14 km (with account of refraction effects). If the scattering is Rayleigh-dominated, than the intensity values will be proportional to the pressure at the altitude $h_B$, it is the basis of Boltzmann temperature estimation.

Contribution of dust causes the effect of depolarization of single scattering. To describe it numerically, Ugolnikov and Maslov (2013b) define the polarization characteristics value $q_0$:

$$q(\theta, h_B) = q_0(h_B)\, q_R(\theta) \qquad (1).$$

Here $q$ and $q_R$ are measured and Rayleigh scattering polarization (second normalized Stokes vector component) depending on angle $\theta$. For most part of twilights the value $q_0$ was close to unity. The minimum of $q_0$ over summer 2012 observations was reached in the evenings of August, 9$^{th}$ and 12$^{th}$. Depolarization effect was also seen during mid-August twilights in 2013. Fig.2 shows the angular dependencies of scattered light polarization $q(\theta)$ at different altitudes $h_B$ in the morning of August, 10$^{th}$ compared with Rayleigh curve. This twilight is characterized by the strongest single scattering depolarization effect in mesosphere over all WAPC observations during three summers.

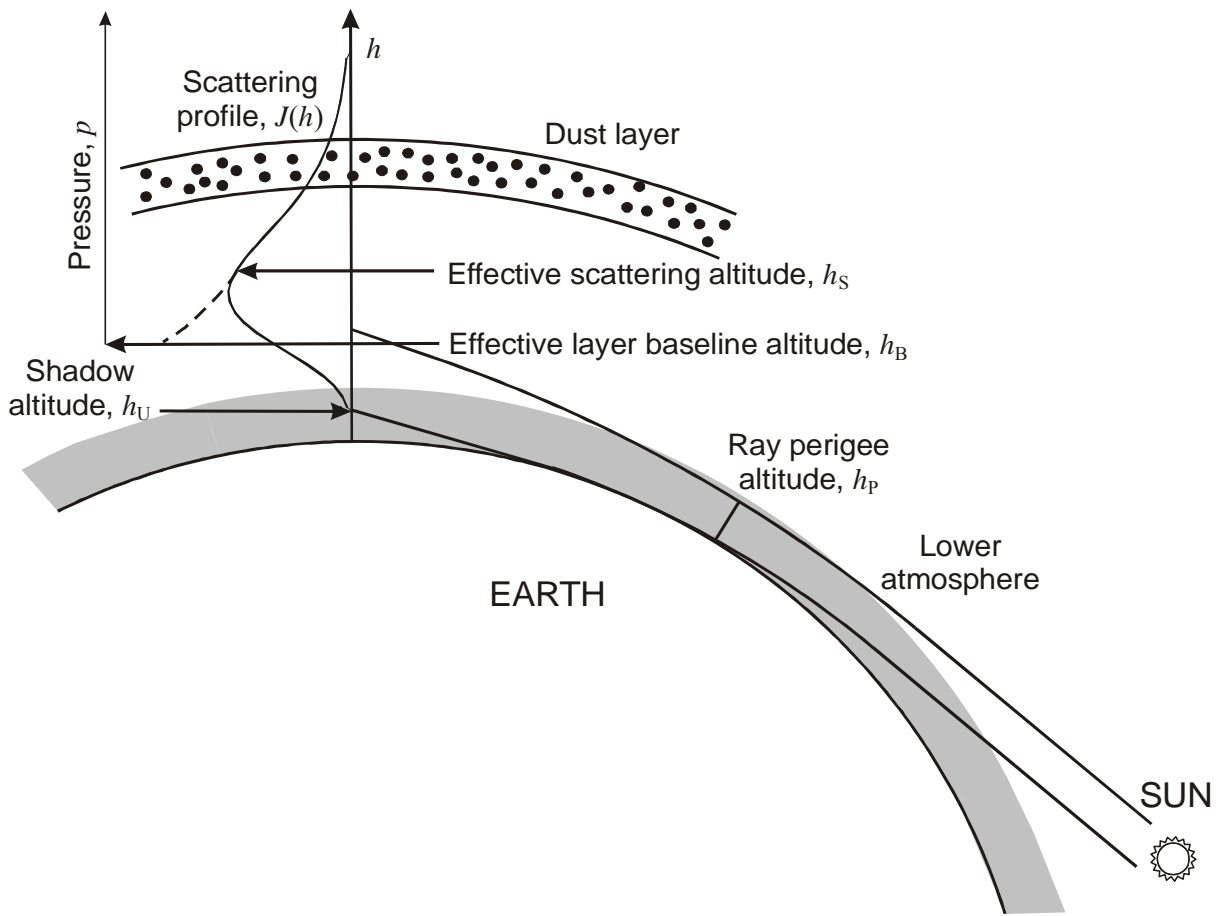

*Figure 1. Geometry of single scattering during the deep twilight.*

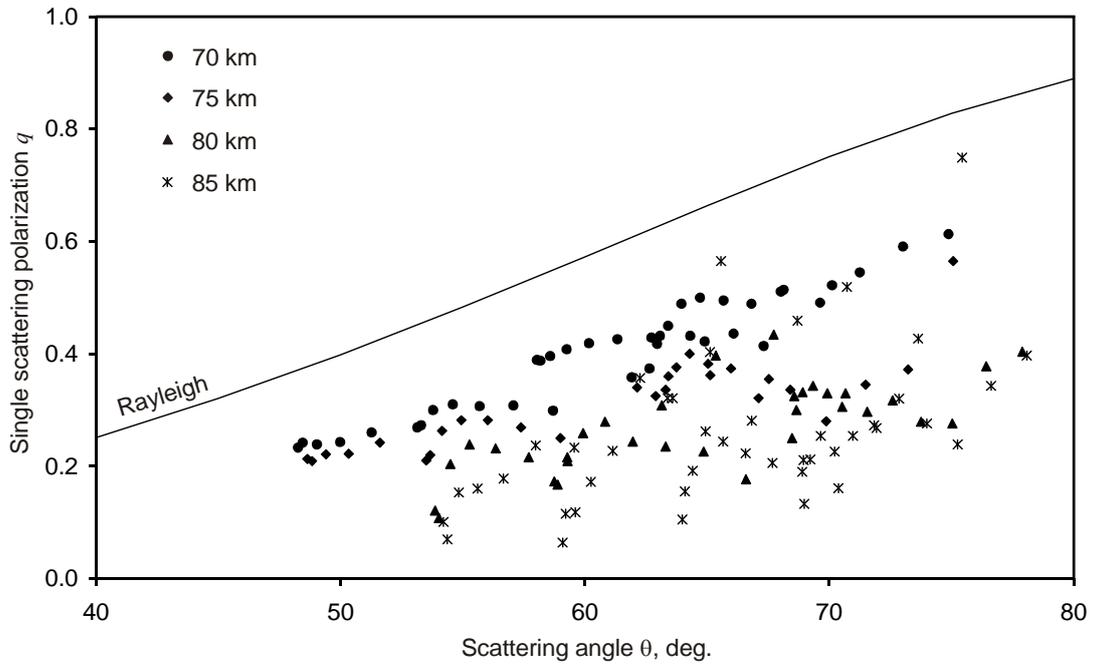

*Figure 2. Angular dependencies of single scattering polarization for different layer baseline altitudes $h_B$ for the morning twilight, August, 10$^{th}$, 2013.*

Fig.3 shows the $q_0$ values at 80 km for all August observations in 2012 and 2013. We see that both evening twilights of August, 9$^{th}$, were characterized by moderate depolarization values. Strong depolarization was expected to appear in the morning, since the Perseids radiant ascends high above the horizon and the amount of dust inflow increases. Evening twilight is characterized by less inflow and, possibly, higher altitude of dust moderation by the tangent trajectory from radiant close to the horizon. More difficult question is the moment of strongest depolarization, preceding the visual maximum of Perseids in 2013 in 2.5 days, and secondary $q_0$ drop following the visual maximum. It will be considered in Discussion chapter.

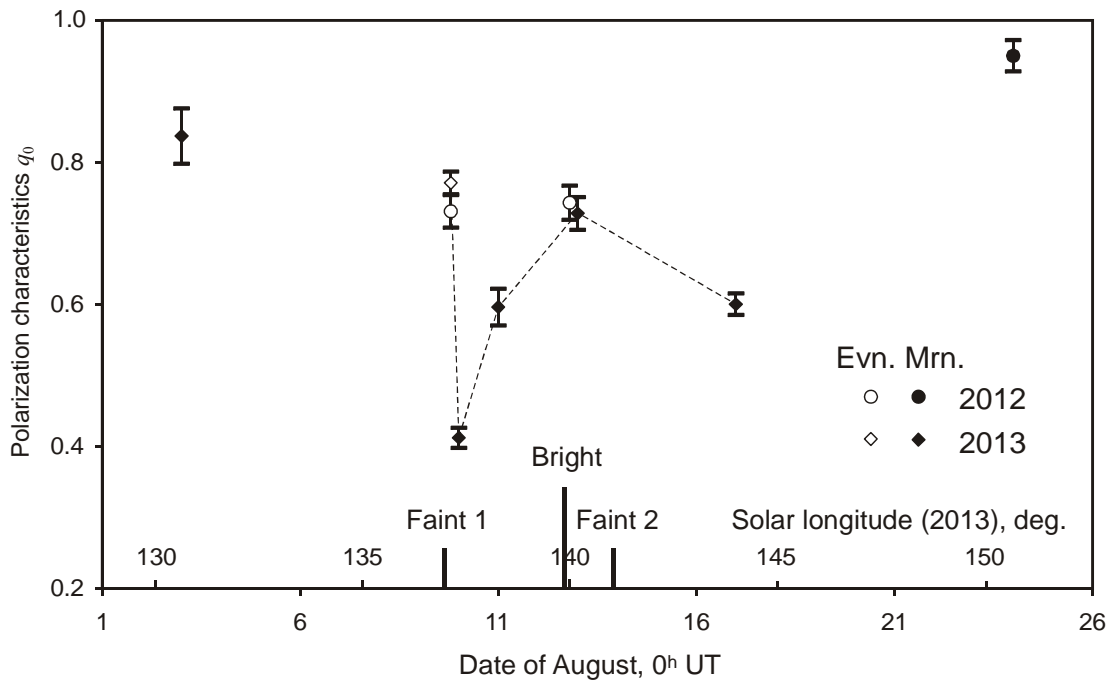

*Figure 3. Polarization characteristics at $h_B$=80 km for August twilights in 2012 and 2013. Positions of bright and faint meteor maxima of Perseids are shown.*

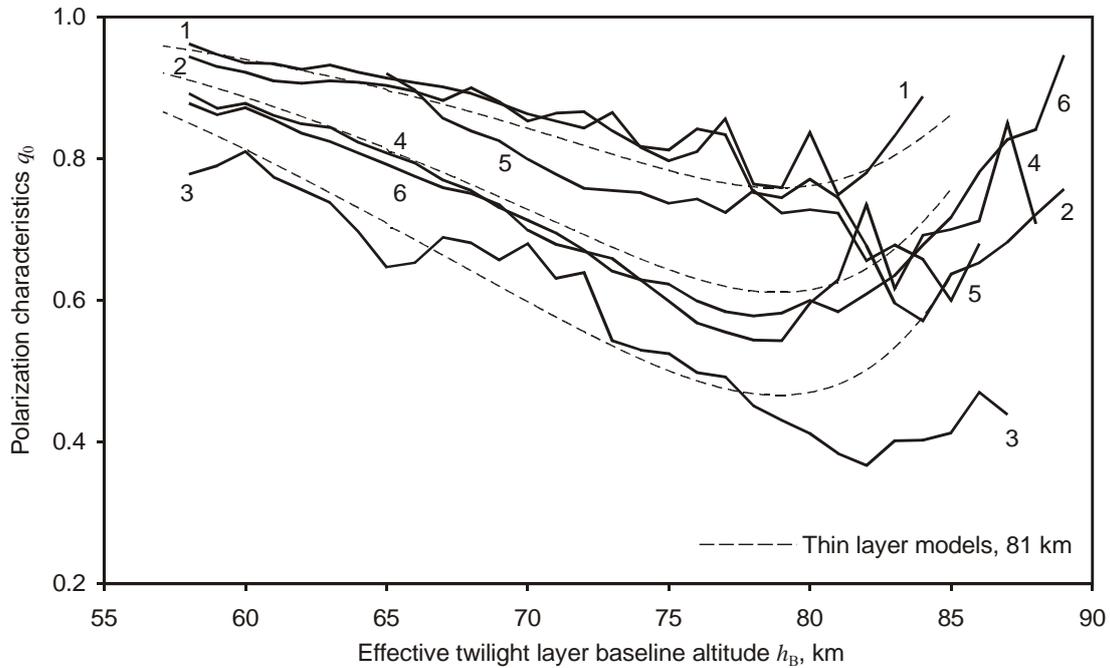

*Figure 4. Altitude dependencies of polarization characteristics for August twilights in 2013 (1 – morning, 3$^{d}$, 2 – evening, 9$^{th}$, 3 – morning, 10$^{th}$, 4 – morning, 11$^{th}$, 5 – morning, 13$^{th}$, 6 – morning, 17$^{th}$). Model curves for thin layer at 81 km (different dust densities) are shown.*

The dependencies of polarization characteristics on layer baseline altitude for August 2013 twilights are shown in Fig.4. It can seem that the dust layer is expanded to the whole mesosphere, since the depolarization effects are seen in a wide range of layer baseline altitudes. In fact, the twilight layer itself is thick enough (see Fig.1), and observable effects can be caused by thin dust layer. Fig.4 also shows the calculated $q_0$ profiles corresponding to such layer at 81 km, those are in good agreement with most part of observation curves. Real thickness of the layer is then hard to estimate. The $q_0(h_B)$ dependencies of pre-maximum twilights (evening, 9$^{th}$, morning, 10$^{th}$ of August) correspond to higher layer altitude, about 83 km. Depolarization is seen until the layer is emitted by the Sun and vanishes at baseline altitudes higher than the layer altitude, when it immerses into the shadow.

Admixture of dust scattering leads to polarization and intensity effects. If we consider the layer baseline altitudes $h_B$ less than dust layer altitudes, than the polarization characteristics $q_0$ will fall with $h_B$ rise due to faster decrease of Rayleigh scattering contribution, the value of $dq_0/dh_B$ will be negative. Total single scattering intensity will decrease with $h_B$ rise slower than by Boltzmann law, and the temperature defined by the method (Ugolnikov and Maslov, 2013b) will be overestimated. Fig.5 shows the diagram of these values ($dq_0/dh_B$ and $T$) for 70 km. EOS Aura/MLS average temperature for this period is shown by the circle (TIMED/SABER operates in Southern hemisphere only in August). The solid lines correspond to the different polarization characteristics of dust scattering $q_D$, defined analogically by the formula (1).

Observational points have significant errors along both axes, offset of evening and morning twilight data is also seen. It makes difficult to define the dust sky polarization exactly. But it can be concluded that non-polarized scattering model fits the observations better, especially for morning twilights when Perseids dust inflow is stronger.

**4. Discussion.**

The basic result of the work is the detection of non-polarized or weakly-polarized admixture to Rayleigh single scattering in the mesosphere starting around August, 10$^{th}$, 2013. The signs of the same effect were also seen in the same dates in 2012 and most probably related with Perseids meteor shower. The problem is the difference of the date noted above and the maximum of the shower observed at August, 12$^{th}$ (in 2013 – at 16$^{h}$ UT, by International Meteor Organization data).

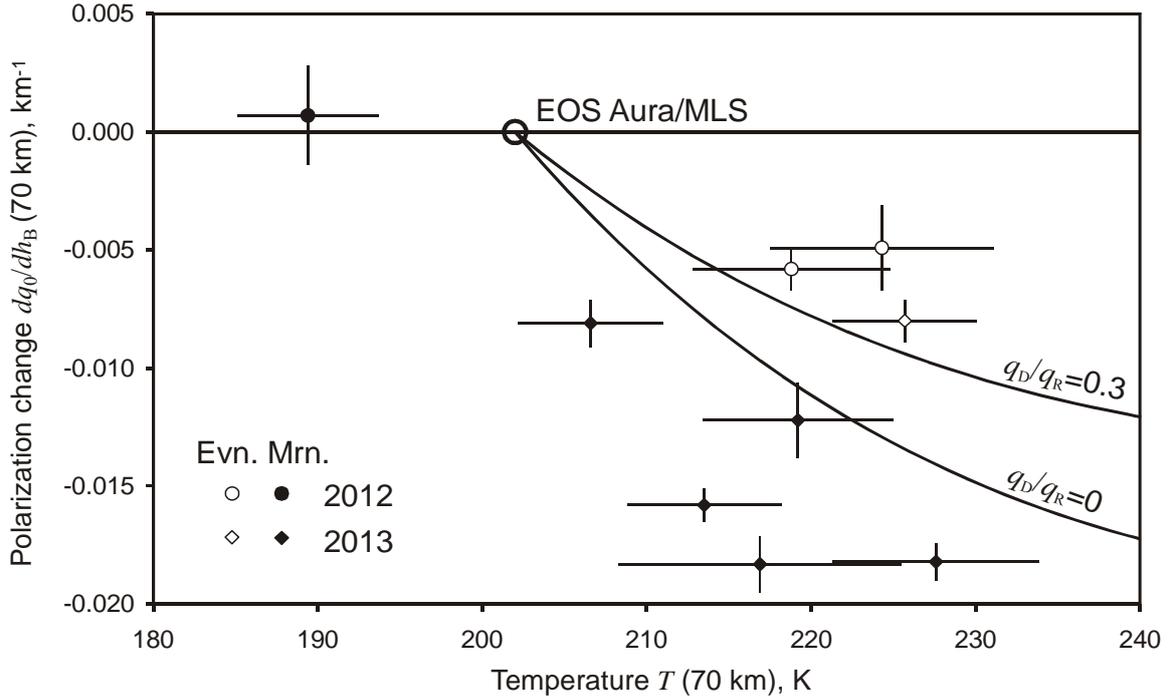

*Figure 5. Polarization change – measured temperature diagram for 70 km in August twilights in 2012 and 2013. Model curves for different dust polarization characteristics are shown.*

The reason of such discrepancy can be related with the structure of Perseids stream and variance of fractions causing the visual meteor shower and the dust moderation in the upper atmosphere. Visual meteor events are created by large particles (size about several millimeters and more). Effective light scattering fraction is formed by smaller particles with sizes about the micron and, possibly, fragmentation products of submillimeter-size particles. Different types of observations – visual (Brown and Rentdel, 1996, Arlt and Buchmann, 2002, and many others) and radar (Šimek and Pecina, 1999) show that Perseids visual maximum at August, $12^{th}$ (solar longitude 140°) is contributed by the bright meteors only. Long-term observational analysis (Bel'kovich and Ishmukhametova, 2006) points out that the profile of low-mass fraction (about $10^{-4}$ g) of Perseids activity has a bimodal structure with two maxima around solar longitudes 137° and 141° (see Fig.3). These maxima become more and more significant for smaller particles. First maximum in 2013 falls to the night of August, 9-$10^{th}$, weakly affecting the evening twilight picture due to low radiant position, but causing the strong $q_0$ drop in the morning. Second low-mass maximum is missed in the observations due to weather conditions but followed by the morning twilight of August, $17^{th}$, also with significant depolarization effect.

The dependencies of single scattering polarization characteristics on twilight layer baseline altitude show that the dust appears in upper mesosphere, basically at the altitudes 81-83 km. These altitudes are also characterized by maximum of NLC density in June and July and close minimum of ozone concentration (Cevolani and Pupillo, 2003), showing the role of meteoric dust and ions in NLC condensation and ozone destruction.

## 5. Conclusion.

All-sky polarization measurements of twilight background give the possibility to retrieve the scattering properties of upper atmosphere and detect the depolarization effect caused by dust. Temperature measurements based on background intensity analysis give the additional information to study the dust scattering polarization. August 2012 and 2013 twilights data show that this value is close to zero meaning the particle sizes higher than the wavelength, but lower than the visual meteors sizes. The numerical parameters obtained in this paper are still uncertain, the problem can be solved by increased number of observations in the future.

**Acknowledgements**

The work is financially supported by Russian Foundation for Basic Research, grant №12-05-00501.